\documentclass[english,aps,prb,showpacs,superscriptaddress,twocolumn]{revtex4-1}
\usepackage[latin9]{inputenc}
\usepackage{textcomp}
\usepackage{amstext}
\usepackage{graphicx}

\makeatletter

\@ifundefined{textcolor}{}
{%
 \definecolor{BLACK}{gray}{0}
 \definecolor{WHITE}{gray}{1}
 \definecolor{RED}{rgb}{1,0,0}
 \definecolor{GREEN}{rgb}{0,1,0}
 \definecolor{BLUE}{rgb}{0,0,1}
 \definecolor{CYAN}{cmyk}{1,0,0,0}
 \definecolor{MAGENTA}{cmyk}{0,1,0,0}
 \definecolor{YELLOW}{cmyk}{0,0,1,0}
}

\usepackage{graphics}\usepackage{epsfig}

\makeatother

\usepackage{babel}
\begin{document}

\title{Knight shift and nuclear spin relaxation in Fe/\emph{n}-GaAs heterostructures}

\author{K. D. Christie}

\affiliation{School of Physics and Astronomy, University of Minnesota, Minneapolis,
MN 55455}

\author{C. C. Geppert}

\affiliation{School of Physics and Astronomy, University of Minnesota, Minneapolis,
MN 55455}

\author{S. J. Patel}

\affiliation{Dept. of Materials, University of California, Santa Barbara, CA 93106}

\author{Q.~O.~Hu}

\affiliation{Dept. of Electrical and Computer Engineering, University of California,
Santa Barbara, CA 93106}

\author{C. J. Palmstr$\o$m}

\affiliation{Dept. of Electrical and Computer Engineering, University of California,
Santa Barbara, CA 93106}

\affiliation{Dept. of Materials, University of California, Santa Barbara, CA 93106}

\author{P. A. Crowell}

\affiliation{School of Physics and Astronomy, University of Minnesota, Minneapolis,
MN 55455}

\email{crowell@physics.umn.edu}

\begin{abstract}
We investigate the dynamically polarized nuclear-spin system in Fe/\emph{n}-GaAs
heterostructures using the response of the electron-spin system to
nuclear magnetic resonance (NMR) in lateral spin-valve devices. The
hyperfine interaction is known to act more strongly on donor-bound
electron states than on those in the conduction band. We provide a
quantitative model of the temperature dependence of the occupation
of donor sites. With this model we calculate the ratios of the hyperfine
and quadrupolar nuclear relaxation rates of each isotope. For all
temperatures measured, quadrupolar relaxation limits the spatial extent
of nuclear spin-polarization to within a Bohr radius of the donor
sites and is directly responsible for the isotope dependence of the
measured NMR signal amplitude. The hyperfine interaction is also responsible
for the $2\text{ kHz}$ Knight shift of the nuclear resonance frequency
that is measured as a function of the electron spin accumulation.
The Knight shift is shown to provide a measurement of the electron
spin-polarization that agrees qualitatively with standard spin transport measurements.
\end{abstract}
\maketitle

\section{Introduction}

Hyperfine interactions profoundly influence electron-spin dynamics
in \emph{n}-GaAs at temperatures below 100~K.\cite{Paget1977,Salis2009,Chan2009,Kolbl2012}
The strong influence is a direct result of the low channel doping
of between $\text{2 - 10}\times10^{16}\text{ cm}^{-3},$ which is
typically used to maximize the electron spin lifetime.\cite{Dzhioev2002,Lou2006,Lou:NP:2007,Chan2009,Kolbl2012}
This doping range is only slightly above the metal-insulator transition
of GaAs, and so the system is best described with a combination of
localized and itinerant electronic states.\cite{Romero1990,Intronati2012}
The wavefunctions of localized electrons have a dramatically enhanced
overlap with nearby nuclei, greatly increasing the efficiency of dynamic
nuclear polarization by the contact hyperfine interaction.\cite{Lampel1968,Abragam1961}
The action of the spin-polarized nuclear system on the electron system
is equivalent to an effective magnetic field that is often larger
than the applied field.\cite{Overhauser1953}

In this paper, we report on measurements of nuclear magnetic resonance
(NMR) by probing the response of the electronic spin accumulation
to this effective field.
A typical means of modulating the electron
spin accumulation is by dephasing the spins with an applied magnetic
field, which is known as the Hanle effect. The presence of the effective
nuclear magnetic field can partially cancel the applied field and
restore (in part) the electron spin polarization. In this experiment, we extend the earlier work of Ref.~\onlinecite{Chan2009} by using NMR to probe the detailed dynamics of the coupled electron-nuclear spin system,
allowing for the extraction of information about the occupancy of donor sites by spin-polarized electrons and their coupling to the nuclear spins of the different isotopes.
When the nuclear
spin-polarization is destroyed by NMR, the electronic spin accumulation
changes in the presence of the new effective magnetic field. Therefore
to detect NMR, we monitor the polarization of the electronic spin
system as a function of the frequency of the applied ac magnetic field.

We show that quadrupolar relaxation of the nuclear spin allows a nonzero
nuclear spin polarization to exist only very near donor sites and
that this spatial dependence explains the order of magnitude difference
in the NMR signal magnitude as a function of isotope. The presence of spin-polarized
electrons acts through the hyperfine interaction as an effective magnetic
field affecting the nuclei near these donor sites. If the
occupation fraction of donor sites is known, the electronic field
near donor sites can be calculated for a given spin accumulation.
We provide a quantitative estimate for the occupation fraction on
the basis of resistivity measurements. Using this model, we show that
the magnitude of the electronic spin polarization can be determined
using the Knight shift.

\begin{figure}[h]
\includegraphics{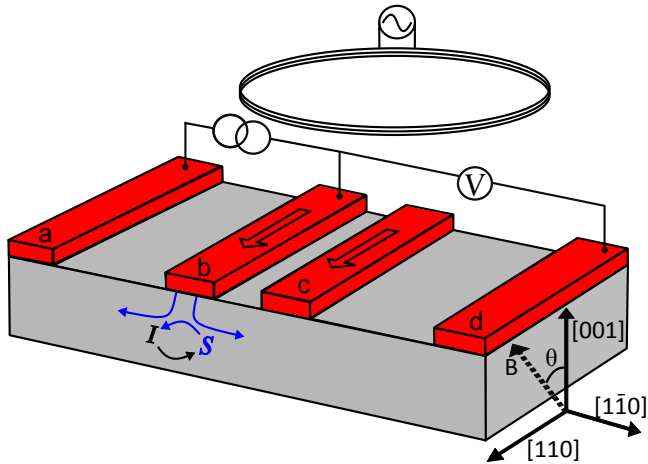}

\protect\caption{\label{Device} (Color online) Schematic of a lateral spin valve device.  A spin accumulation $S$ is generated in the
GaAs channel in a three-terminal configuration and detected as a change
in voltage $\Delta V_{bd}$.  The coil placed over the device is used as a source of ac magnetic field for the NMR measurements.  The nuclear spin $I$ is coupled to $S$ through the hyperfine interaction, as represented by the two arrows.}
\end{figure}

Figure~\ref{Device} shows a schematic of a typical lateral spin-valve
device. From bottom to top, the epitaxial Fe/\emph{n}-GaAs (100) heterostructures consist of a GaAs buffer layer followed by a Si-doped 2.5~$\mu$m thick channel  $(n \sim 3-8\times10^{16}\text{ cm}^{-3})$, a 15 nm $n\rightarrow n^+$ transition layer over which the Si-doping is increased to $5\times 10^{18}$~cm$^{-3}$, followed by a 15 nm thick $n^+$ ($5\times10^{18}$~cm$^{-3}$) layer.\cite{Hanbicki2003}  The Fe layer is 5~nm thick and is grown at a nominal substrate temperature of $0^\circ$~C.\cite{Lou:NP:2007,Lou2006,Chan2010}
The structures are capped with thin layers of Al and Au. The heterostructures
are fabricated using standard photolithography and semiconductor processing
techniques into lateral spin-valves with injection and
detection contacts (5 $\mu$m~\texttimes ~50 $\mu$m), labeled $b$
and $c$ respectively, separated by 10 $\mu$m. The heavily doped interfacial regions form Schottky tunnel barriers.  A spin-polarized current
is created at the injection contact ($b$) when the Fe/GaAs interface
is biased. This spin current leads to a non-equilibrium spin accumulation
$S$ in the channel, where
\begin{equation}
S=\frac{1}{2}\frac{n_{\uparrow}-n_{\downarrow}}{n_{\uparrow}+n_{\downarrow}},
\end{equation}
and $n_{\uparrow(\downarrow)}$ is the concentration of electrons
with spin up(down). The spin accumulation is established in the channel
by the combined effects of drift, diffusion, relaxation and precession.
The presence of the spin accumulation in the channel is detected as
a change in voltage relative to a remote contact $d$, at either the
injection contact itself in a three-terminal configuration $\Delta V_{bd}$,\cite{Lou2006,Tran2009,Chan2009}
or at a nonlocal detection contact $\Delta V_{cd}$ (connection not
shown).\cite{Lou:NP:2007,Salis2010} In either case, the spin accumulation
can be detected by dephasing the spins in the channel using the Hanle
effect with an applied magnetic field perpendicular to the magnetization
of the contacts.

\begin{figure}[bh]
\includegraphics{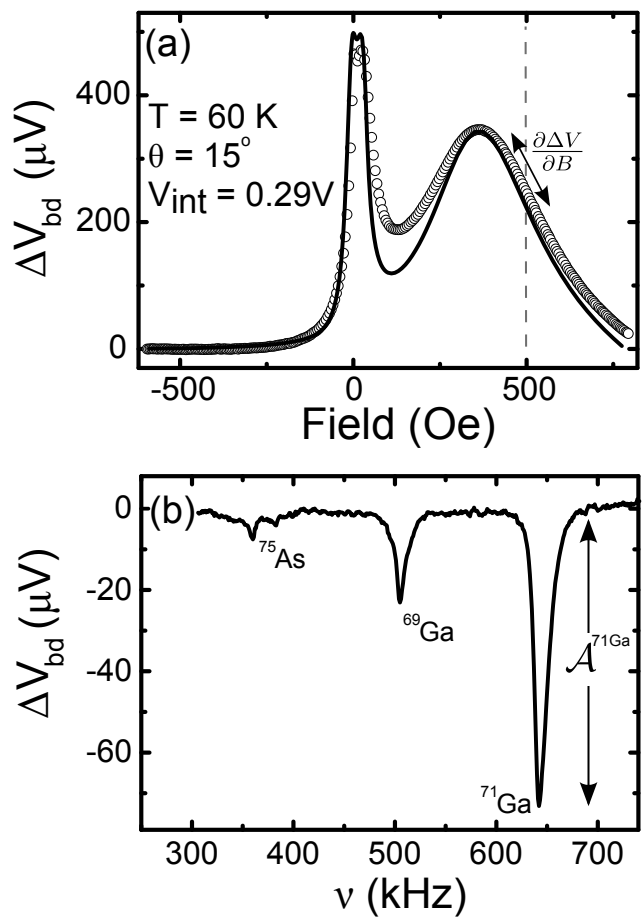}\protect\caption{\label{Hanle-NMR}(a) The change in the three-terminal Hanle voltage
$\Delta V_{bd}$ as a function of magnetic field (open circles) for Fe/\emph{n}-GaAs ($5\times10^{16}\text{ cm}^{-3}$).  A fit to the corresponding
spin diffusion model including the nuclear field as described in Ref.~\onlinecite{Chan2009} is shown as a solid curve.
The slope of the spin signal at large field $\partial\Delta V_{bd}/\partial B$
determines the sensitivity of the electronic spin accumulation to
NMR. (b) The change in the electronic spin signal $V_{bd}$ at 500
Oe as a function of the frequency of the AC magnetic field. Three
NMR peaks are observed corresponding to the three isotopes of GaAs.
The magnitude of the NMR signal $\mathcal{A}$ is observed to be an
order of magnitude larger for $^{71}\text{Ga}$ than for $^{75}\text{As}$.}
\end{figure}

Figure \ref{Hanle-NMR}(a) shows the change in three-terminal voltage
$\Delta V_{bd}$ as a function of the applied field $B$. When the
field is oriented at an oblique angle $\theta=15^{\circ}$ from the
sample normal as depicted in Fig.~\ref{Device}, two peaks are discernible
in the Hanle signal as a function of the applied field. The peak at
non-zero field ($\approx 375$~Oe in Fig.~\ref{Hanle-NMR}) is a result
of the cancellation of the applied field by the effective nuclear
field.\cite{Paget1977,Meier:1984} To measure NMR, an ac magnetic
field is applied using a few-turn coil placed above the sample as shown
in Fig.~\ref{Device}. The nuclear field is reduced when the frequency of the
ac field matches the nuclear resonance condition $\nu^{\alpha}=\gamma^{\alpha}B_{a}$,
where $\gamma^{\alpha}$ is the nuclear gyromagnetic ratio for the
nuclear isotope $\alpha$, and $B_{a}$ is the magnitude of the applied
field. The slope of the spin signal at a fixed applied field $\frac{\partial\Delta V}{\partial B}$
determines the sensitivity to a change in nuclear field. For example
in Fig. \ref{Hanle-NMR}(a), the slope is negative and large at 500
Oe. Figure \ref{Hanle-NMR}(b) shows the change in the electron spin
signal $V_{bd}$ as a function of the ac frequency with the static field
fixed at 500 Oe; the decrease in the spin signal at the resonance
has an amplitude $\mathcal{A}^{\alpha}$ for each isotope $\alpha$.
Note that the measured amplitude at the resonance of $\alpha={}^{71}\text{Ga}$
is an order of magnitude larger than at the resonance of $^{75}\text{As}$.
Similar differences in the relative magnitudes of NMR signals from
different isotopes have been observed at temperatures above 50 K in other dynamically polarized
samples doped between 2 and 10$\times10^{16}\text{ cm}^{-3}$ as well
as in $\text{Al}_{1-y}\text{Ga}_{y}\text{As}$ systems.\cite{Chan2009,VanDorpe2005}

In the following sections we show that the observed isotope dependence
of the magnitude of the NMR signal arises from the fact that the hyperfine
and quadrupolar nuclear-spin relaxation rates are of comparable magnitude.
In Section \ref{hyperfine} we review the model of a coupled electron-nuclear spin
system. In Section \ref{Gamma} we demonstrate a simple means of determining
the temperature dependence of the donor occupation fraction, which
determines the efficiency of DNP, from charge transport. In Section
\ref{Quad} we show that the temperature dependence of the measured NMR signal
can be reproduced with a quantitative model accounting for the spatial
distribution of spin-polarized nuclei. The different NMR signals for
each isotope are shown to be a result of a different effective volume
of spin-polarized nuclei around donors. In Section \ref{Knight} we show that
it is possible to measure the Knight shift of the nuclear resonance
frequency by using spin transport. We use the Knight shift to extract
an alternative measurement of the spin-polarization of the electron system.

\section{the coupled electron-nuclear spin system}\label{hyperfine}

Just above the metal-insulator transition, the bottom of the GaAs
conduction band can be described as a combination of localized (impurity) and
itinerant states. Electron-electron interactions effectively maintain
the same average spin polarization between these states.\cite{Paget1981}
The electrons in localized states interact strongly with the lattice
nuclei via the contact hyperfine interaction
\begin{equation}
H=\frac{8\pi}{3}g_{e}\mu_{B}\gamma_{N}\mathbf{I}\cdot\mathbf{S}\left|\psi_{e}\right|^{2},
\end{equation}
where $\mu_{B}$ is the Bohr magneton, $g_{e}$ is the free electron
g-factor, $\gamma_{N}$ is the nuclear gyromagnetic ratio, $\mathbf{I}$
and $\mathbf{S}$ are the nuclear and electronic spin operators respectively,
and $\psi_{e}$ is the electronic wavefunction evaluated at a nuclear
site. For the purposes of our model, we assume the localized wavefunction
is that of a hydrogenic donor-bound electron
\begin{equation}
\psi_{e}=\frac{1}{\sqrt{\pi a_{o}^{3}}}e^{-r/a_{o}}
\end{equation}
with an effective Bohr radius $a_{o}=\text{10 nm}$.\cite{Shklovskii1984}
Assuming a hydrogenic wavefunction, Paget \emph{et~al.\cite{Paget1977}}
recast the hyperfine interaction in terms of two effective magnetic
fields: the Knight field, an effective electronic field acting on
the nuclei that depends on the distance from the donor:
\begin{equation}
B_{e}=\Gamma b_{e}\mathbf{S}e^{-2r/a_{o}},
\label{eq:Be}
\end{equation}
and the Overhauser field, an effective nuclear field acting on the
electrons that is the weighted sum of all the nuclei in the electron\textquoteright s
Bohr radius:
\begin{equation}
B_{N}=fb_{N}\mathbf{I}.
\label{eq:Bn}
\end{equation}
The strength of the electronic and nuclear fields  $b_{e}$ and $b_{N}$
have been calculated for GaAs to be\cite{Paget1977,Meier:1984}
\begin{equation}
b_{e}=-170\text{ G\ and \ }b_{N}=-53\text{ kG.}
\end{equation}
The occupation factor $\Gamma$ and leakage factor $f$ take into
account that
$B_{e}$ and $B_{N}$ are smaller than their maximum values.
$\Gamma$ represents the fraction of donors that are occupied by an electron.  Only these donors can contribute to $B_e$.
The Overhauser field $B_N$ is reduced
from its maximum possible value by the leakage factor $f$, which
takes into account relaxation of the nuclear system by all other channels besides the hyperfine interaction.

The leakage factor
can be easily motivated by considering the following rate equation
for a nuclei with spin $I=3/2:$\cite{Abragam1961}
\begin{equation}
\frac{d\mathbf{I}}{dt}=\frac{4}{3}I(I+1)\frac{\mathbf{S}}{T_{H}}-\frac{\mathbf{I}}{T_{H}}-\frac{\mathbf{I}}{T_{1}^{*}},
\end{equation}
where the first term represents the polarization of nuclei by spin-polarized
electrons, and the second and third terms represent hyperfine relaxation
with a rate $T_{H}^{-1}$ and all other nuclear spin relaxation mechanisms
at a rate $T_{1}^{*-1}$ respectively. In the steady-state limit ($dI/dt = 0$), the
average nuclear spin $\mathbf{I}$ is proportional to the average
electron spin polarization $\mathbf{S}$ and the ratio of the pumping
rate $T_{H}^{-1}$ due to hyperfine coupling to the total nuclear
relaxation rate $T_{1}^{-1}=T_{H}^{-1}+T_{1}^{*-1}$:
\begin{equation}
\mathbf{I}=\frac{4}{3}I(I+1)\mathbf{S}T_{H}^{-1}/(T_{H}^{-1}+T^{*-1}).
\end{equation}
The leakage factor $f$ is ratio of the hyperfine to total relaxation
rate and can be written as
\begin{equation}
f=\frac{T_{1}^{*}}{T_{H}+T_{1}^{*}},\label{eq:Leakage}
\end{equation}
so that
\begin{equation}
\mathbf{I}=f\frac{4}{3}I(I+1)\mathbf{S}.
\end{equation}
When the hyperfine relaxation time $T_{H}$ is small, $f$ is unity
and the nuclear spin polarization is maximized. If, however, other
channels for nuclear spin relaxation besides hyperfine coupling are
present, then $f<1$ and the polarization is reduced from its ideal
value.

The fields $\Gamma b_{e}$ and $fb_{N}$ can
be determined from our data by modeling the coupled electron-nuclear spin dynamics.\cite{Paget1977,Chan2009}
In these models the average Overhauser field is\cite{Paget1977,Dyakonvo1973}
\begin{equation}
\vec{B}_{N}=fb_{N}\frac{4}{3}(I+1)\frac{(\vec{B}+\Gamma b_{e}\vec{S})\cdot\vec{S}}{(\vec{B}+\Gamma b_{e}\vec{S})^{2}+B_{o}^{2}}(\vec{B}+\Gamma b_{e}\vec{S}),\label{eq:OverhauserField}
\end{equation}
where on average the nuclear spins are oriented along the vector sum of the
applied field $\vec{B}$ and electronic field $\Gamma b_{e}\vec{S}$. The
dipolar field (sometimes called the local field) is described with
a phenomenological constant $B_{o}$.\cite{Paget1977,Gammon2001,Abragam1961}
Chan \emph{et~al.\cite{Chan2009}} determined the electronic and nuclear
fields as well as $B_{o}$ in an Fe/\emph{n}-GaAs device with channel
doping of 5$\times 10^{16} $ cm$^{-3}$ by numerically solving the
drift-diffusion equations for the electronic spin accumulation self-consistently
in the presence of the nuclear field given in Eq. \ref{eq:OverhauserField} and fitting to Hanle curves measured in an oblique magnetic field.
They determined the nuclear fields to be
\begin{equation}
\Gamma b_{e}=-50\,\text{G}\ \ \text{and}\ \ fb_{N}=-15\,\text{kG}
\end{equation}
at 60 K as well as measuring the effective dipolar field to be $B_{o}=50\text{ G}$.
Comparing to the theoretically calculated values, the implied occupation
factor was $\Gamma=0.3$. In the following section we show that this
value agrees with estimates determined from charge transport.

\section{Occupation Fraction\label{sec:Occupation-Fraction}}\label{Gamma}

As discussed in the previous section, the fractional occupation $\Gamma$
of the localized states is a parameter which can be measured by modeling
the coupled electronic and nuclear spin dynamics. In this section
we provide a means of extracting $\Gamma$ from a model of the electrical
resistivity that takes into account the conduction and impurity bands.
This model is independent of spin transport measurements and can be
used to predict the value of the nuclear hyperfine relaxation rate as
well as the Knight field.

\begin{figure}
\includegraphics{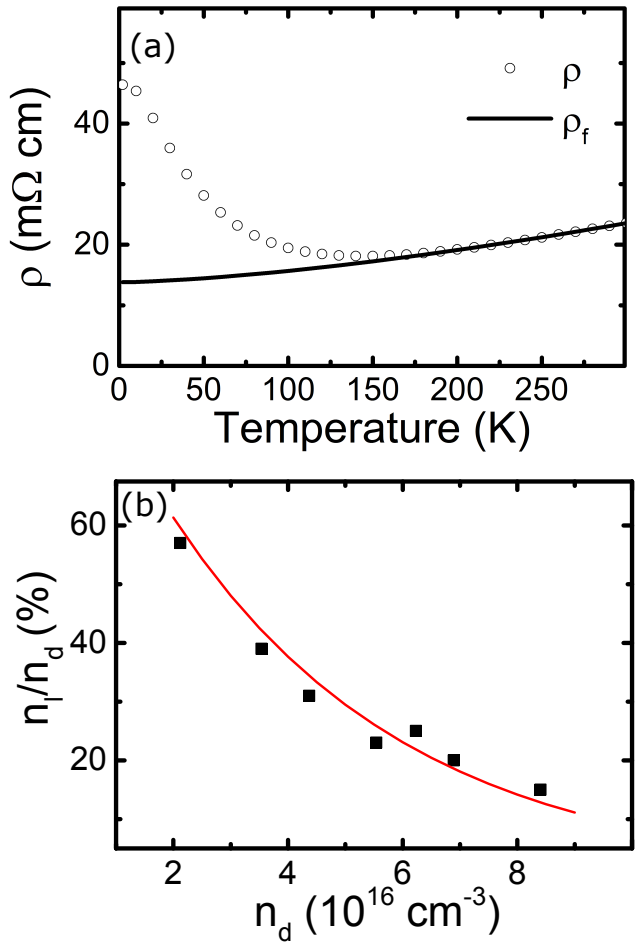}\protect\caption{\label{FigResistivity} (Color online) (a) The resistivity $\rho$ (open symbols) of the GaAs channel
as a function of temperature for a sample with $n_{d}$= 5$\times 10^{16}$ cm$^{-3}$ (open circles). A fit to the contribution $\rho_{f}$, which includes phonon and ionized impurity scattering, is shown as the black line. The difference
$\rho(T)-\rho_f(T)$ can be used to calculate the occupation fraction of donors $\Gamma(T)$
as described in the text. (b) The fraction of localized
states $n_{L}/n_{d}$ (symbols) as a function of doping concentration $n_{d}$,
assuming that $\Gamma(T=0)=1$ as described in the text. The red solid
line is a fit based on the Hertz distribution with a critical radius
$r_{c}=18\text{ nm}$ (see text).}
\end{figure}

Figure \ref{FigResistivity}(a) shows the resistivity as a function
of temperature for a sample doped at 5$\times 10^{16}$ cm$^{-3}$.
At 300~K, all of the donors are ionized and the resistivity can be
treated as being metallic: $\rho_{f}=\rho_{I}+\rho_{OP}=\rho_I+cT^{3/2}$,
where $\rho_{I}$ takes into account ionized impurity scattering in
the degenerate limit, and the term with a pre-factor $c$ takes
into account the phonon contribution to the resistivity.\cite{Szmyd1990,Jacoboni2010,Lancefield1987}
A curve representing $\rho_f$ is shown in Fig.~\ref{FigResistivity}(a) as a solid black line.
As the temperature is lowered, the resistivity drops until electrons
become localized by occupying states in the impurity band, at which point the resistivity
increases dramatically. We attribute the increase in resistivity to
a combination of the decrease in the number of itinerant electrons as they freeze
out into localized sites as well as a decreased electron mobility
in the localized impurity band.   Using Matthiessen's rule we model
the resistivity
\begin{equation}
\label{eq:rho1}
\rho=\frac{m}{ne^2\tau}=\frac{m}{e^2}\left (\frac{1}{n_{d}-n_{L}\Gamma(T)}\right )\left (\tau_I^{-1} +\tau_{OP}^{-1}+\tau_{IB}^{-1}\right ),
\end{equation}
where the three scattering rates are due to ionized impurity scattering, optical phonon scattering, and scattering in the impurity band (neutral impurity scattering), and the density of carriers is reduced from the donor density $n_d$ by the number density $n_L\Gamma(T)$ of occupied isolated donors.   Equation~\ref{eq:rho1} can be recast in terms of resistivities:
\begin{equation}
\label{eq:Resistivity}
\rho(T)=\frac{1}{1-\frac{n_L\Gamma(T)}{n_d}}\left [\rho_f(T) +\rho_{IB}\frac{n_L\Gamma(T)}{n_d}\right ],
\end{equation}
where $\rho_{IB}$ is the impurity band resistivity in the limit in which all donors are neutralized.  The actual impurity band contribution to the resistivity at any temperature is $\rho_{IB}$ multiplied by the fraction $\Gamma(T)n_L/n_d$ of singly occupied (neutral) and isolated donors.
$\rho_{IB}$ has been investigated in several
semiconductor systems including Ge,\cite{Fritzsche1955,Nishimura1965,Davis1965}
and GaAs.\cite{Woodbury1973,Agrinskaya2010} Our system corresponds
to the intermediate doping range in which conduction occurs in the
upper-Hubbard or $D^{-}$ band. This band is composed of
donor sites occupied by 2 electrons ($D^{-}$ states).\cite{Shklovskii1984} Based on the increase in resistance
at low temperatures in previously studied Ge and GaAs samples,\cite{Davis1965,Woodbury1973}
as well as simple models of an electron scattering off of neutral
donors,\cite{Erginsoy1950,Honig1966,Maxwell1966,Kozhevnikov1995}
the resistivity $\rho_{IB}$ due to localized impurity states is estimated to be 5-10 times larger than that in the conduction band
at low temperatures.  We have found that using values of $\rho_{IB}$
between $5\rho_{I}$ and $10\rho_{I}$ only varies the final value
of $\Gamma$ in the analysis below by less than 20\%. For the
data shown in Fig.~\ref{FigResistivity}(a), we use an intermediate
value
\begin{equation}
\rho_{IB}=7.5\rho_{I}=100\,\text{m\ensuremath{\Omega}cm}
\end{equation}
to determine the contribution $\rho_{IB}$ of the localized impurity states to the resistivity.
Eq. \ref{eq:Resistivity} can then be solved for the occupation
factor
\begin{equation}
\Gamma(T)=\frac{n_d}{n_L}\left [\frac{\rho(T)-\rho_f(T)}{\rho(T)+\rho_{IB}}\right ].\label{eq:GammaFromResistivity}
\end{equation}

Assuming a given value of $\rho_{IB}$, the only other unknown parameter in Eq.~\ref{eq:GammaFromResistivity} required in order to determine $\Gamma(T)$ is
the number of localized states $n_{L}$, which can be estimated from
simple statistics. We consider a donor site to be localized if the
closest neighboring donor is further away than a critical distance $r_{c}$.
The number of localized sites is given by the Hertz distribution:\cite{Hertz1909}
\begin{equation}
n_{L}=n_{d}\intop_{r_{c}}^{\infty}dr\,\frac{3}{a_{d}}\left(\frac{r}{a_{d}}\right)^{2}e^{-(r/a_{d})^{3}},\label{eq:Hertz}
\end{equation}
where $a_{d}$ is the average distance between donor sites $a_{d}=\left(\frac{4\pi n_{d}}{3}\right)^{-1/3}.$
The number of localized states can be estimated experimentally
as a function of doping. Figure \ref{FigResistivity}(b) shows the
value $n_{L}/n_{d}$  for
samples doped between $3-8\times10^{16}\text{ cm}^{-3}$, estimated by assuming the occupation factor $\Gamma$ is
unity in the limit of zero temperature. These data
can be fit with Eq. \ref{eq:Hertz}, and a single value of $r_{c}=18\text{ nm}$
is found to reproduce $n_{L}/n_{d}$ over this doping range. The estimated
critical radius $r_{c}$ is approximately twice the Bohr
radius $a_{o}=10\text{ nm}$, which suggests that the localized states
considered here can still be treated as hydrogenic donors.

\begin{figure}
\includegraphics{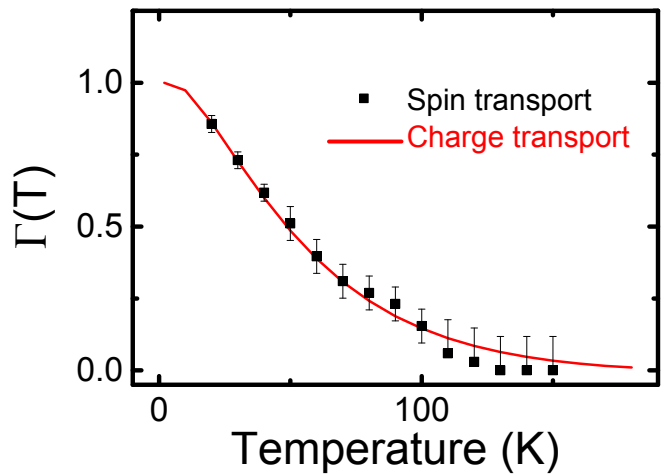}

\protect\caption{\label{fig:Gamma} (Color online) The occupation fraction $\Gamma$ of donors as a function
of temperature. As discussed in the text, the occupation fraction
can be determined by modeling spin-transport in the presence of nuclear
spins (symbols), or can it can be determined from the resistivity by using Eq.
\ref{eq:GammaFromResistivity} (solid curve).}

\end{figure}

The solid curve in Fig.~\ref{fig:Gamma} shows the occupation factor $\Gamma$
as a function of temperature calculated from the resistivity as described above.
The points represent the occupation factor as a function of temperature
taken from the measurements of Ref.~\onlinecite{Chan2010}, where $\Gamma b_{e}$ is determined
from spin transport by fitting the coupled electron-nuclear spin system
in the same way as described by Chan \emph{et al}.\cite{Chan2009}
In this case, the occupation factor is calculated by assuming that
the majority of the Knight field arises from Ga nuclei. In the next
section we validate this assumption by showing that spin-polarized
As nuclei account for only 1\% of the total Overhauser field at 60K.
The two measurements of $\Gamma$ agree to within experimental uncertainties.
This result shows that the electronic field comes from localized electronic
states distinct from the conduction band. It also provides a
simpler means of estimating the hyperfine relaxation rate rather than
modeling the electron spin dynamics at each temperature.

\section{The nuclear leakage factor}\label{Quad}

The data of Fig. \ref{Hanle-NMR} show an order of magnitude difference
in the magnitude of the measured NMR signals of $^{71}\text{Ga}$
and of $^{75}\text{As}$ at 60 K. We now show that this difference
arises from a combination of the spatial dependence of the hyperfine
relaxation rate and the strong isotope dependence of nuclear quadrupolar
relaxation rate.  In this model, the leakage factor becomes a function of position.  The hyperfine relaxation rate can be estimated as
\begin{equation}
\frac{1}{T_{H}^{\alpha}}=\tau_{c}\Gamma(T)(b_{e}^{\alpha}\gamma_{N}^{\alpha})^{2}e^{-4r/a_{o}},
\label{eq:hrate}
\end{equation}
where $\tau_{c}$ is the correlation time of the hyperfine interaction,
$b_{e}^{\alpha}$ is the electronic field acting on the nuclear isotope
$\alpha$, and $\gamma_{N}^{\alpha}$ is the nuclear gyromagnetic
ratio for that isotope.\cite{Paget1977}  Physically, $\tau_c$ is the time scale over which the hyperfine field fluctuates due to the relaxation and subsequent repolarization of the
spin-polarized electron bound to the donor.  When the repolarization process is efficient, as we expect for electrical spin injection,
we expect $\tau_c$ to be of the same order as the spin relaxation time $\tau_s$.  Essentially, Eq.~\ref{eq:hrate} describes a relaxation process in which the nuclear polarization is dephased by precession in
the fluctuating hyperfine field, which has a root mean-square value of $b_e^\alpha e^{-2r/a_0}.$   As in a motional narrowing process, a shorter correlation time (more rapid fluctuation), results in a smaller relaxation rate.
The hyperfine relaxation rate becomes exponentially
smaller for nuclei further from the donor site.

It has been shown that in undoped GaAs, where hyperfine coupling is irrelevant, that
Raman-like scattering of phonons dominates nuclear spin relaxation
at temperatures above 30 K.\cite{McNeil1976} We therefore equate $T_{1}^{*-1}$, the nuclear relaxation rate due to non-hyperfine processes, with the quadrupolar relaxation
rate $T_{Q}^{-1}$ in our calculation of the leakage factor (Eq.  \ref{eq:Leakage}). The electric
quadrupole moment of the nuclei couples to the phonons via the electric
field gradient induced by these scattering events. The quadrupolar
Hamiltonian is non-spin-conserving, resulting in a decrease in the
average nuclear spin as phonons are excited. The resulting quadrupolar
relaxation rate is given as\cite{McNeil1976}
\begin{equation}
\frac{1}{T_{Q}^{\alpha}}=\kappa(Q^{\alpha}T)^{2},\label{eq:Qrate}
\end{equation}
where $\kappa$ is is a parameter which takes into account the coupling
between phonons and the nuclei, $Q^{\alpha}$ is the quadrupole moment,
and $T$ is the temperature. The parameter $\kappa$ is expected to
be independent of doping and of nuclear isotope and is treated as
a fitting parameter. The values of the nuclear quadrupolar moments
for the isotopes in GaAs are,\cite{Lide2000}
\begin{eqnarray}
Q^{75\text{As}} & = & 314\,\text{mb,}\\
Q^{69\text{Ga}} & = & 171\,\text{mb},\\
Q^{71\text{Ga}} & = & 107\,\text{mb},
\end{eqnarray}
where 1~mb$=10^{-31}$~m$^{-2}$.  At fixed temperature, the quadrupolar relaxation rate
for $^{75}\text{As}$ should therefore be 9 times larger than for $^{71}\text{Ga}$.

In our system, the hyperfine and quadrupolar relaxation mechanisms
are similar in magnitude near donor sites. We consider the case of
nuclei around a donor occupied by single spin-polarized electron.
The leakage factor is
\begin{equation}
f^{\alpha}(r)=\frac{T_{Q}^{\alpha}}{T_{Q}^{\alpha}+T_{H}^{\alpha}}=\frac{1}{1+\frac{\kappa}{\tau_{c}}\left(Q^{\alpha}T\right)^{2}/\Gamma(T)\left(b_{e}^{\alpha}\gamma_{N}^{\alpha}\right)^{2}e^{-4r/a_{o}}},
\end{equation}
which is explicitly isotope and position dependent. The leakage factor
$f$ is maximized near the donor where the hyperfine coupling is strongest,
and $f=0$ as $r\rightarrow\infty$. Therefore the nuclear field is
only large in the region around donor sites where the hyperfine interaction
dominates.

\begin{figure}
\includegraphics{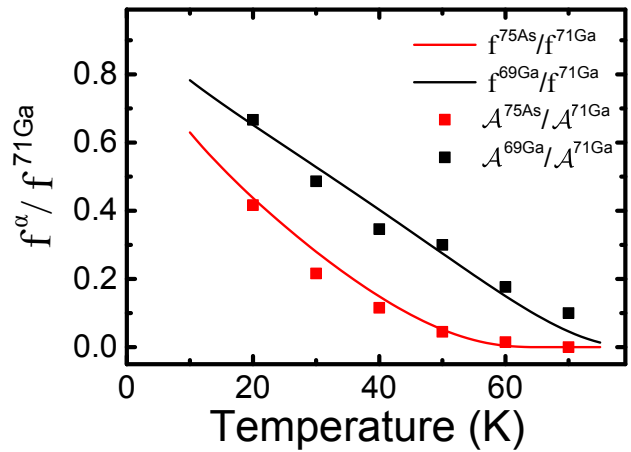}\protect\caption{\label{FigureLeakageFactors} (Color online) The ratio of the leakage factor $f$
of $^{75}\text{As}$ (red line) and $^{69}\text{Ga}$ (black line) to that of $^{71}\text{Ga}$
as a function of temperature. As discussed in the text, the ratio
of the leakage factors is predicted to be the same as the ratio $\mathcal{A}^{\alpha}/\mathcal{A}^{\text{71Ga}}$
of the magnitude of the NMR signals  (symbols). By using the value $\kappa/\tau_{c}=20\frac{\text{Hz}^{2}}{\text{mb}^{2}\text{K}^{2}}$
in the calculation of the quadrupolar radius (Eq. \ref{eq:QuadrupolarRadius})
the measured NMR signal magnitudes are reproduced for all temperatures.}
\end{figure}

We model the spatial extent of the polarized nuclear spins by defining
an effective quadrupolar radius $r_{Q}$ at which the two relaxation rates $T_{H}^{-1}$ and
$T_{Q}^{-1}$ are equal:\cite{Paget2008}
\begin{equation}
r_{Q}=-\frac{a_{o}}{4}\ln\left[\frac{\kappa}{\tau_{c}\Gamma(T)}\left(\frac{Q^{\alpha}T}{\gamma_{N}^{\alpha}b_{e}^{\alpha}}\right)^{2}\right].
\label{eq:rq}
\end{equation}
For $r>r_{Q}$, the hyperfine
relaxation rate decreases exponentially, and to a good approximation
the leakage factor can be assumed to be zero. The effective
leakage factor is therefore given by the weighted average
\begin{equation}
f^{\alpha}=\frac{4}{a_{o}^{3}}\intop_{0}^{\infty}r^{2}dr\, f^{\alpha}(r)e^{-2r/a_{o}}\simeq\frac{4}{a_{o}^{3}}\intop_{0}^{r_{Q}}r^{2}dr\, f^{\alpha}(r)e^{-2r/a_{o}}.\label{eq:QuadrupolarRadius}
\end{equation}
The value of the NMR signal $\mathcal{A^{\alpha}}$ is proportional
to the following quantities: the sensitivity of the spin signal with respect to the
applied field $\frac{dV}{dB}$, the induced change in nuclear polarization,
and the isotope dependent leakage factor $f^{\alpha}$. This yields
the following relation:
\begin{equation}
\mathcal{A}^{\alpha}\propto\frac{dV}{dB}f^{\alpha}b_{N}\Delta I.
\end{equation}
Therefore the ratio of measured resonance signals of any two isotopes
at any temperature should be equal to the ratio of their leakage factors.
Figure \ref{FigureLeakageFactors} shows the ratio of the magnitude
of the NMR signal of $^{75}\text{As}$ and $^{69}\text{Ga}$ to that
of $^{71}\text{Ga}$ as a function of temperature. The lines are the
calculated ratios of the leakage factor taken from our model as a
function of temperature. We are able to model the complete temperature
dependence by assuming the ratio $\kappa/\tau_{c}$ is constant. We
have repeated this analysis on several samples in the doping range
of 2 - $8\times10^{16}\text{cm}^{-3}$ and found in all cases that
\begin{equation}
\kappa/\tau_{c}=20\pm2\frac{\text{Hz}^{2}}{\text{mb}^{2}\text{K}^{2}}
\end{equation}
 reproduces the temperature dependence of the NMR signals. This result
agrees with our previous assumption that quadrupolar relaxation in
samples doped just above the metal-insulator transition is independent
of doping. It also shows that quadrupolar relaxation essentially localizes
the nuclear polarization within a Bohr radius of each donor site.
The order of magnitude difference in the measured electronic response
to NMR among different isotopes is a direct result of the spatial
extent of this nuclear polarization.  We emphasize that the strong coupling between electron and nuclear spins implies that the strong spatial inhomogeneity in the nuclear polarization should impact electron spin transport and dynamics.

\section{Knight Shift}\label{Knight}

Because the hyperfine field exists only around an occupied donor,
the argument of the previous section implies that the nuclear polarization
is largest around donors. The spatial average of this field for all
three isotopes determines the average hyperfine field experienced by the
electrons. The fact that this field decreases rapidly with increasing
temperature is due to the $T^{2}$ dependence in Eq. \ref{eq:Qrate}.  In comparison,~$\Gamma(T)$, which governs the hyperfine relaxation rate, depends  more weakly on temperature over the range of this experiment ($< 60$~K).  At higher temperatures, $\Gamma(T)$ decreases exponentially, leading to an even stronger temperature dependence of the hyperfine field.
We now shift our focus to the measurement of the electronic field
acting on the nuclei by measuring the Knight shift of the resonance
frequency as a function of the electronic spin accumulation.

Resonance occurs at the nuclear Larmor frequency $\nu^{\alpha}=\gamma_{N}^{\alpha}B_{tot}$,
where the total field $\vec{B}_{tot}$ is the vector sum of the applied field $\vec{B}_{a}$
and the much smaller Knight field $\vec{B}_{e}^\alpha(S)$ acting on isotope $\alpha$ due to the electronic spin polarization $S$. Note that $\vec{B}_e$ is parallel to $\vec{S}$.  In the limit where the Knight field
is much smaller than the applied field, the nuclear resonance frequency
takes the form
\begin{equation}
\nu^{\alpha}=\gamma_N^{\alpha}\left|\vec{B_{a}}+\vec{B}_{e}^\alpha(S)\right|\approx\gamma_{N}^{\alpha}B_{a}+\gamma_{N}^{\alpha}B_{e}^\alpha(S)\sin\theta+\mathcal{O}(\frac{B_{e}^{2}}{2 B_{a}}),
\end{equation}
where $\theta$ is the oblique angle of the applied field indicated
in Fig. \ref{Device}.  The Knight shift is hence determined by the component of $\vec{S}$ parallel to $\vec{B}_a$.  At a fixed applied field, the Knight shift
of the resonance frequency $\Delta\nu^{\alpha}$ is directly proportional
to spin polarization via the electronic field $B_e^\alpha(r)$:
\begin{equation}
\Delta\nu^{\alpha}= \gamma_N^{\alpha} \Gamma(T) \langle B_{e}^\alpha(r)\rangle S \sin\theta\,\label{eq:KnightShift}
\end{equation}
where $\langle B_e^(r)\rangle$ is the electronic field averaged over the envelope of the donor-bound electron wave-function.  We assume that the polarized nuclei are confined to a sphere of radius $r_Q$ around the donor, where $r_Q$ is the quadrupolar radius, so that
\begin{equation}
\langle B_e^\alpha(r)\rangle = b_e^\alpha \frac{\int_{0}^{r_Q} r^2 e^{-2r/a_0}dr}{\int_{0}^{r_Q} r^2 dr},
\end{equation}
where, as found in Ref.~\onlinecite{Paget1977},
\begin{equation}
b_{e}^{\text{As}}=-220\,\text{G,}\ \ b_{e}^{\text{Ga}}=-130\,\text{G}.
\end{equation}
At 60~K, the average electronic fields using the quadrupole radii calculated from Eq.~\ref{eq:rq} are -194~G for $^{75}$As and -83~G for $^{69}$Ga.

\begin{figure*}[t]
\includegraphics{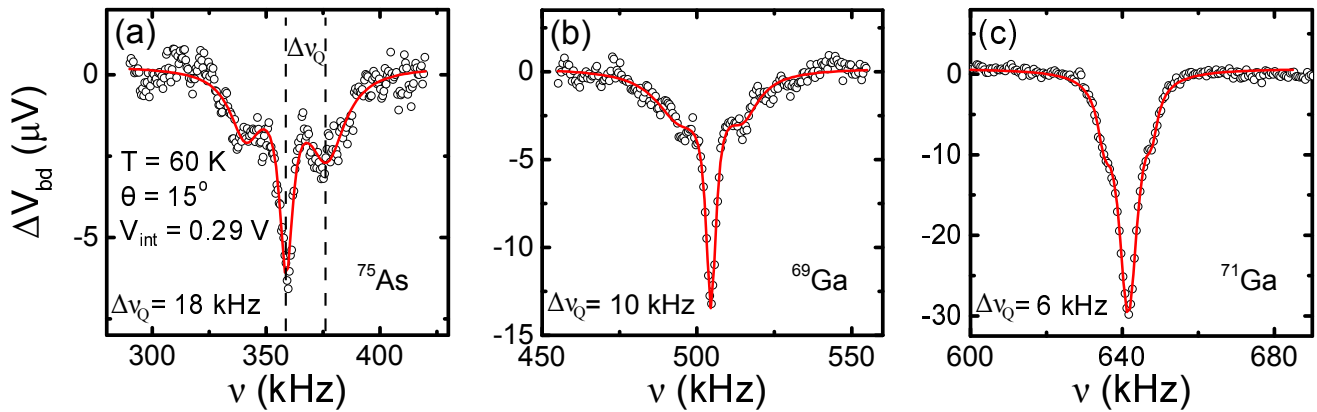}
\protect\caption{\label{FigureDetailedNMR}(Color online) The change in the spin-signal $\Delta V_{bd}$
as a function of the ac magnetic field frequency for (a) $^{75}\text{As}$,
(b) $^{69}\text{Ga}$, and (c) $^{71}\text{Ga}$. The static field is 490~Oe. A detailed frequency
scan reveals a triple peak structure of each isotope, which is attributed
to strain induced quadrupolar splitting of the nuclear Zeeman transitions.
Fits assuming three Lorentzians are shown as solid lines for each
isotope. The measured splittings of the side peaks from the central
peak ($\Delta\nu_{Q}$) are the quadrupolar splitting in a uniaxial
electric field gradient. }
\end{figure*}

Equation \ref{eq:KnightShift} predicts the Knight shift of the resonance
frequency to be approximately twice as large for $^{75}\text{As}$
than for $^{69}\text{Ga}$: $\Delta\nu^{^{75}\text{As}}\approx2\text{ kHz}$
for $60$ K and $\theta=30^{\circ}$. A high precision measurement
of the resonance frequency is therefore required to observe the Knight
shift. Figure \ref{FigureDetailedNMR} shows the change in the electron spin accumulation as a function of frequency taken
using a very slow frequency sweep rate (14 Hz/s) to ensure that the nuclear
system remains in steady state. Each resonance has three peaks
as a result of a crystal strain field interacting with the quadrupole
moment of an isotope with spin $I=3/2$.\cite{Abragam1961,Guerrier1997}
We have verified for each isotope and for several angles that
the satellite peaks have a difference in frequency $\Delta\nu_{Q}$
relative to the central peak that is in agreement with the standard
formula for quadrupolar splitting in an single uniaxial electric field
gradient $V_{zz}$ oriented along the direction perpendicular to the
Fe/GaAs interface:
\begin{equation}
h\Delta\nu_{Q}^{\alpha}=\frac{eV_{zz}Q^{\alpha}}{4I(2I-1)}\frac{3\cos^{2}\theta-1}{2}\left[3m_{z}^{2}-I(I+1)\right],
\end{equation}
where $Q^{\alpha}$ is the electric quadrupole moment of isotope $\alpha$,
$m_{z}$ is the magnetic quantum number, and $I=3/2$ is the nuclear
spin.\cite{Slichter1990} Fits of the resonance curves assuming a
triple Lorentzian model are shown as solid lines in Fig. \ref{FigureDetailedNMR}. The Knight shift is determined from the frequency of the central
peak measured as a function of the bias current, which determines
the electron spin polarization.

\begin{figure}
\includegraphics{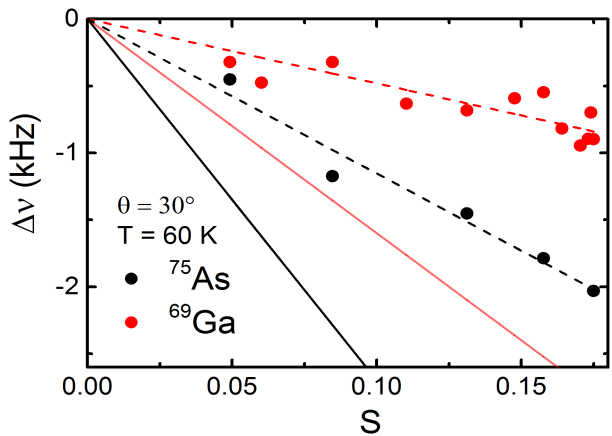}

\protect\caption{\label{FigureKnightShift} (Color online) Isotope dependent Knight shift (symbols)
as a function of the spin accumulation calculated from spin transport.  The dashed lines are linear fits of the data.
Using the occupation factor calculated from the resistivity, the Knight
shift is calculated exactly using Eq. \ref{eq:KnightShift}, and the result is shown for each isotope using the solid lines.}
\end{figure}

The Knight shift is measured at an angle of $\theta=30^{\circ}$.
This nearly doubles the frequency shift relative to the data in Fig.
\ref{FigureDetailedNMR}. Figure \ref{FigureKnightShift} shows the
resonance frequency taken from fits of NMR curves of $^{75}\text{As}$
and $^{69}\text{Ga}$ as a function of the electron spin $S$ measured
by spin transport. The Knight shift was not measured for $^{71}\text{Ga}$
because at high angles $\Delta\nu_{Q}^{\text{71Ga}}$ becomes smaller
than the NMR line width, making an accurate fit of the resonance frequency
impossible.

In an all-electrical spin transport experiment, the spin polarization
is directly proportional to the magnitude of the non-local spin signal
$\Delta V_{cd}:$
\begin{equation}
S=\frac{e\Delta V_{cd}}{\eta}\frac{g(\epsilon_{f})}{n},\label{eq:spinaccumulation}
\end{equation}
where $\eta$ is the spin detection efficiency of the Fe/GaAs interface
at zero bias and $g(\epsilon_{f})/n$ is the density of states at the Fermi level
normalized by the carrier density in the GaAs channel.\cite{Hu2011,Salis2009,Salis2010}
For Fe/GaAs interfaces the detection efficiency has been measured
to be $\eta\approx0.2$ in spin-LED's.\cite{Adelmann2005} The lines
shown in Fig. \ref{FigureKnightShift} are the expected Knight shifts
determined from Eq. \ref{eq:KnightShift}, with the non-local electron spin
polarization determined from Eq. \ref{eq:spinaccumulation} and extrapolating the measured polarization back to the injector.

As can be seen in Fig.~\ref{FigureKnightShift}, the absolute magnitudes of the Knight shifts predicted by Eq.~\ref{eq:KnightShift} are larger than the experimental values by a factor of $\approx 3$.  The  measured ratio $\Delta \nu (^{75}\text{As})/\Delta \nu (^{69}\text{Ga})$ of the shifts is approximately 2.4 in experiment, while the expected ratio is 1.7.
Given the limitations of the experiment, we do not believe that these discrepancies are that significant.  The absolute electron spin polarization $S$ is impacted by
uncertainties in $\eta$ as well as the fact that the density-of-states used in applying Eq.~\ref{eq:spinaccumulation} is assumed to be that of an ordinary parabolic conduction band with an effective mass $m^*= 0.07 m_e$.  We estimate that systematic errors in $S$ are of the order of 50\%.  In any case, the maximum value of $S$ inferred from Eq.~\ref{eq:spinaccumulation} is $\approx 0.2$, which corresponds to a polarization $P\approx 0.4$,  which we consider an \emph{upper bound} given the the known spin polarization of iron.  Any reduction in  $S$ would improve the agreement in Fig.~\ref{FigureKnightShift}, and we conclude that the value extracted from transport is probably too large by a factor of approximately two.  As expected, the Knight shift for $^{69}\text{Ga}$ is smaller than for $^{75}\text{As}$.  The ratio of the shifts for the two isotopes is very sensitive to the values of the quadrupole radii $r_Q$ calculated from Eq.~\ref{eq:rq}, which was based on several assumptions.  This would easily account for the discrepancy in the observed ratio.  In summary, we consider the qualitative agreement in Fig.~\ref{FigureKnightShift}, including the sign of the Knight shift, the linearity with $S$, and the relative magnitudes of the shifts for the two isotopes, to be satisfactory.

\section{Summary}

We have provided a quantitative description of NMR in Fe/\emph{n}-GaAs
lateral spin valve devices by exploiting the strong hyperfine coupling
at these dopings. We have shown that the occupation fraction $\Gamma$
of donors can be estimated from charge transport. The competition
between hyperfine coupling and the quadrupolar nuclear relaxation rate
leads to a spatially inhomogeneous nuclear polarization that is strongest
near donors. The magnitude of the NMR signal of each isotope is directly
proportional to the effective volume of polarized nuclear spins around
donor sites. We also showed that within this volume the nuclei are
directly affected by the presence of the Knight field. Finally, we
measured the Knight shifts of the nuclear resonance frequencies as
a function of the spin-accumulation. Using the calculated occupation
factor, we show that the Knight shift is proportional the spin accumulation as measured by spin transport.

This work was supported by NSF Grants No. DMR-1104951,
the NSF MRSEC Program under DMR DMR-0819885, and by C-SPIN, one of
the six centers of STARnet, a Semiconductor Research Corporation program
sponsored by MARCO and DARPA.

\bibliographystyle{apsrev4-1}
\bibliography{NMRpaper-Christie}

\end{document}